# Physics-informed Transformer Model for the Design of Wavelength-filtering Ring Resonator

Y.D. Lim[1], F.S. Wan[1], R.J. Wan[2] and C. S. Tan[1]

1. School of Electrical and Electronics Engineering, Nanyang Technological University, 639798, Singapore
2. Department of Computer Science, Hong Kong Baptist University, Hong Kong
*yudian.lim@ntu.edu.sg (Email address of corresponding author)*

*Abstract*—We have developed a physics-informed transformer model to suggest design parameters in wavelength-filtering ring resonator, that suit a given pair of resonant wavelengths with <6 nm errors. The model provides a versatile method for rapid and accurate design of resonators corresponding to various resonant wavelengths.

## I. INTRODUCTION

Photonics integrated circuits (PIC) have been widely used in ion trap quantum computing devices for photon-sensing and optical addressing of trapped ion qubits [1], as illustrated in Fig. 1(a). In a typical quantum computing operation, 4 – 5 types of laser lights with various wavelengths are needed, as shown in the inset of Fig. 1(b). For further miniaturization of PIC-integrated ion trap, our research group has introduced a technological disclosure [2], which involves the usage of ring resonator for wavelength filtering purposes. The principle of wavelength-filtering ring resonator for its application in $^{88}Sr^+$ ion traps is illustrated in Fig. 1(b).

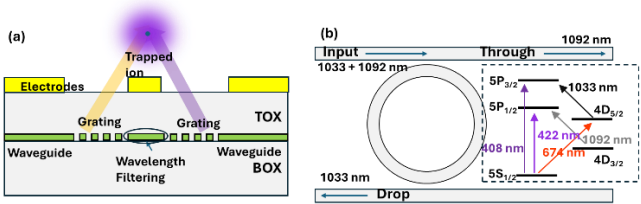

Fig. 1. (a) Illustration of optical addressing using integrated PIC, (b) Ring resonator for wavelength filtering (inset: energy levels in $^{88}Sr^+$ ion)

In the context of $^{88}Sr^+$ ion, wavelength filtering can be achieved by tuning the design of the ring resonator to achieve resonant wavelengths of 1033 and 1092 nm (ref. Fig. 1(b)). The resonant wavelengths can be calculated using equation (1):

$$m \cdot \lambda_{res} = n_{eff} \cdot L \qquad (1)$$

where m, $n_{eff}$, are resonant modes and effective index, respectively. L is the round trip length for the ring resonator, where L = 2πr (r = radius of the ring resonator). From equation (1), multiple factors ($n_{eff}$, L) determine the resonant wavelength. To compute the resonant condition for a single wavelength is straightforward, however, to obtain the ring resonator design for two specified wavelengths may be complicated. Moreover, for every different wavelength pair selected, the fine-tuning of the ring resonator design needs to be carried out from scratch. To resolve this issue, we propose to develop a machine learning (ML) model which uses the desired resonant wavelengths as inputs to suggest the corresponding resonator design. ($n_{eff}$ and r).

## II. PHYSICS-INFORMED TRANSFORMER MODEL

As mentioned earlier, we attempt to develop an ML model for rapid suggestion of $n_{eff}$ and r values to achieve specified resonant wavelengths. Transformer model is selected, as the multi-head attention mechanism in the model can compute the correlation each resonant wavelength. For the training data, we computed >30,000 rows of wavelengths/$n_{eff}$/r datasets using $n_{eff}$ from 1.8 to 2.2 (corresponds to the possible $n_{eff}$ of SiN waveguide) and r values from 1.5 to 5 µm. Resonant wavelengths range between 1,000 to 1,100 are recorded as the model inputs. The transformer model can accept various numbers of input wavelengths from 2 to 7. Among the >30,000 rows of datasets, 10% of the data are used as the testing data (not used for model training).

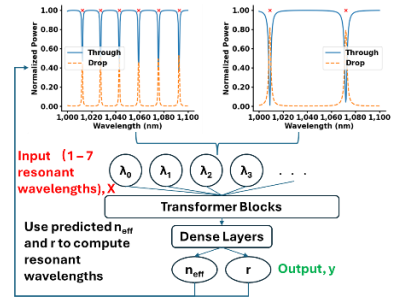

Fig. 2 Illustration of the transformer model

In typical ML model training, the model constantly adjusts the parameters in the model, in order to minimize its loss. Typically, the loss can be computed using equation (2):

$$loss = \frac{1}{n}\sum(y_{pred} - y_{true})^2 \qquad (2)$$

where $y_{pred}$ and $y_{true}$ represent the predicted and actual $n_{eff}$/r values, respectively. For such model training, it did not consider the physics aspect of the relationship between resonant wavelengths and the $n_{eff}$/r values. In this work, we propose a custom loss function improvised from the previously-reported physics-informed neural network [3]. The loss function is:

$$\text{custom loss} = \text{loss} + \alpha \cdot \frac{1}{n}\sum(X_{computed} - X_{true})^2 \quad (3)$$

where $X_{computed}$ are resonant wavelengths computed using the prediction $n_{eff}$/r values ($y_{pred}$) and α is a constant factor. By using the custom loss function, we incorporated the physics consideration when training the transformer model.

## III. RESULTS AND DISCUSSION

As referred to Fig. 2 and equation (3), we attempted to use models with 1 – 4 transformer blocks and various α values, representing the weights of physics component in model training. After training the models, the isolated 10% dataset are used to test the model. In this work, instead of comparing the prediction/actual $n_{eff}$ and r values, we used the prediction/actual $n_{eff}$/r values to compute the resonant wavelengths of the >3,000 rows testing dataset. Then, the median absolute errors between resonant wavelengths computed by predicted/actual $n_{eff}$ and r values are compiled, as shown in Fig. *3*(a). In general, transformer models trained using custom loss (equation (3)) show lower absolute error compared to models trained using conventional loss function. From Fig. *3*(a), transformer model with 4 transformer blocks and α = 0.05 shows lowest error. For further investigations, we computed the absolute errors of all rows in the testing data, as shown in Fig. *3*(b). The absolute errors of the computed resonant wavelengths mostly fall between 1.4 – 6.6 nm.

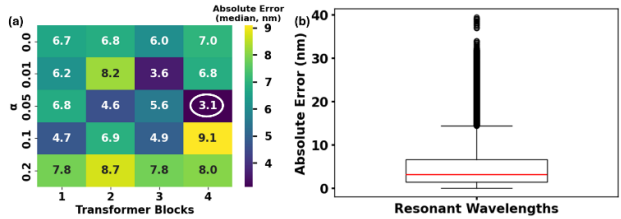

Fig. 3. (a) Median absolute errors of computed resonant wavelengths from transformer models, (b) Absolute errors of resonant wavelengths computed by neff and r values predicted using model with 4 transformer blocks and α= 0.05

In the context of wavelength filtering for [88]Sr[+] ion qubits, resonant wavelengths should occur at the wavelengths correspond to the energy levels illustrated in Fig. 1(b). To test the trained model (4 transformer blocks, α= 0.05) for such application, we attempted to use the trained model to suggest suitable $n_{eff}$ and r values for the filtering of 1,033 and 1,092 nm. As the model accepts 2 – 7 resonant wavelengths with fixed free spectral range (FSR), the input values could be [1033, 1092], [1033, 1062.5, 1092], [1033, 1054.67, 1072.34, 1092]...etc. Preliminary, we attempted to use [1033, 1092] and [1033, 1044.8, 1056.6, 1068.4, 1080.2, 1092]. The suggested values are [r = 1.598 µm, $n_{eff}$ = 1.949] and [r = 4.889 µm, $n_{eff}$ = 2.023], respectively. By using the abovementioned r and $n_{eff}$ pairs, we computed their respective power spectrum in through and drop ports (ref. Fig. 1(b)), as shown in Fig. *4*(a) and (b), respectively.

From Fig. *4*(a), the resonant wavelengths obtained are 1029 and 1087 nm. Benchmarking against the desired 1033 and 1092 nm, the absolute error is 4 – 5 nm. From Fig. *4*(b), the obtained resonant wavelengths are 1002, 1019, 1036, 1053, 1071, and 1090 nm. Benchmarking against the desired 1033 and 1092 nm, the obtained 1036 and 1090 nm shows absolute errors of 2 – 3 nm. As referred to Fig. 1(b) and Fig. *4*, 1033 nm and 1092 nm can be channelled into the input port, where 1033 and 1092 shall be channelled to drop and through ports, respectively.

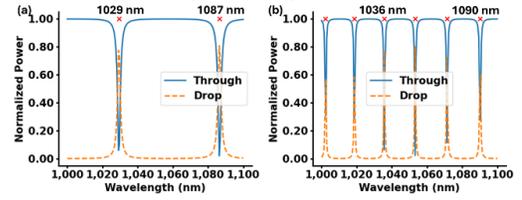

Fig. 4. Power spectrum in through and drop ports from ring resonators with (a) neff = 1.949, r = 1.598 µm, and (b) neff = 2.023, r = 4.889 µm

## IV. CONCLUDING REMARKS

From the previous section, we have demonstrated that the trained transformer model is able to suggest $n_{eff}$ and r values for the filtering of 1033 and 1092 nm. The obtained absolute errors of the resonant wavelengths are 2 – 3 nm and 4 – 5 nm, depending on the input towards the model. In actual ring resonator design, these errors can be easily offset by tuning the phase of the ring. Similar method can be applied to design ring resonator to filter different wavelengths, such as 408, 422 and 674 nm. To enable further expansion of this work, we have included the full Python code for data preparation, model construction, model training, and model evaluation in ref. [4]

ACKNOWLEDGMENT

This work is supported by Ministry of Education of Singapore AcRF Tier 2 (T2EP50121-0002 (MOE-000180-01)) and AcRF Tier 1 (RG135/23, RT3/23); National Semiconductor Translation and Innovation Centre (NSTIC (M24W1NS007)); National Centre for Advanced Integrated Photonics (NCAIP) (NRF-MSG-2023-0002).